\documentclass[11pt,a4paper]{article}

\usepackage[utf8]{inputenc}
\usepackage[english]{babel}
\usepackage{amsmath, amssymb, amsthm}
\usepackage{graphicx}
\usepackage{hyperref}
\usepackage{geometry}
\usepackage[authoryear,round]{natbib}
\usepackage{booktabs}
\usepackage{caption}
\usepackage{subcaption} 
\usepackage{orcidlink}
\usepackage{algorithm}
\usepackage{algorithmicx}
\usepackage{algpseudocode} 

\title{\textbf{Challenges in Applying Variational Quantum Algorithms to Dynamic Satellite Network Routing}}

\author{
  Phuc Hao Do$^{1}$ \\
  \texttt{do.hf@sut.ru,\orcidlink{0000-0003-0645-0021}} \\
  \and
  Tran Duc Le$^{2}$ \\
  \texttt{let@uwstout.edu,\orcidlink{0000-0003-3735-0314}} \\
}

\date{
  $^{1}$Department of Telecommunication Engineering, Bonch-Bruevich St. Petersburg State University of Telecommunications \\
  $^{2}$Mathematics, Statistics and Computer Science, University of Wisconsin–Stout \\[2ex]
}

\begin{document}

\maketitle

\begin{abstract}
Applying near-term variational quantum algorithms to the problem of dynamic satellite network routing represents a promising direction for quantum computing. In this work, we provide a critical evaluation of two major approaches: static quantum optimizers such as the Variational Quantum Eigensolver (VQE) and the Quantum Approximate Optimization Algorithm (QAOA) for offline route computation, and Quantum Reinforcement Learning (QRL) methods for online decision-making. Using ideal, noise-free simulations, we find that these algorithms face significant challenges. Specifically, static optimizers are unable to solve even a classically easy 4-node shortest path problem due to the complexity of the optimization landscape. Likewise, a basic QRL agent based on policy gradient methods fails to learn a useful routing strategy in a dynamic 8-node environment and performs no better than random actions. These negative findings highlight key obstacles that must be addressed before quantum algorithms can offer real advantages in communication networks. We discuss the underlying causes of these limitations, including barren plateaus and learning instability, and suggest future research directions to overcome them.
\end{abstract}


\section{Introduction}
\label{sec:introduction}

The advent of large-scale Low Earth Orbit (LEO) satellite constellations, spearheaded by initiatives such as SpaceX's Starlink, Amazon's Project Kuiper, and OneWeb, is poised to revolutionize global connectivity \cite{saeed2021cubesat}. By deploying thousands of interconnected satellites, these networks promise to deliver high-speed, low-latency internet access to every corner of the globe, including remote and underserved regions \cite{Reddy2023}. However, the very characteristics that enable this new paradigm - namely, the massive scale and high orbital velocity of the satellites - introduce unprecedented challenges in network management \cite{Hu2023}. The network topology is in a constant state of flux, with inter-satellite links (ISLs) being established and terminated on a timescale of seconds, creating a highly dynamic and complex operational environment \cite{Bhattacharjee_2024}.

At the heart of managing these constellations lies the network routing problem: determining the optimal path for data packets to travel from a source to a destination \cite{Zhang2025GRLR,Chen2021Analysis}. In this dynamic context, the routing problem is far more complex than in terrestrial networks. It must account for time-varying latencies, intermittent link availability, and vast state spaces. Formally, finding the optimal path in such a dynamic graph, especially when subject to constraints like ensuring a simple (loop-free) path, is an NP-hard combinatorial optimization problem \cite{garey1979computers}. Classical routing protocols often rely on heuristics that may provide fast but suboptimal solutions, while exact algorithms are computationally intractable for real-time decision-making in networks of this scale.

Quantum computing offers a fundamentally new paradigm for tackling such computationally hard problems \cite{Younis2024}. By leveraging quantum-mechanical principles like superposition and entanglement, quantum computers can, in principle, explore a vast solution space more efficiently than their classical counterparts \cite{raisuddin2024review}. While fault-tolerant quantum computers are still a distant prospect, the current Noisy Intermediate-Scale Quantum (NISQ) era has seen the emergence of hybrid quantum-classical algorithms \cite{preskill2018quantum}. These algorithms are specifically designed to run on near-term, noisy quantum processors. They utilize a classical computer to guide the quantum computation via a variational feedback loop, making them promising candidates for achieving a practical quantum advantage on optimization tasks.

This paper investigates the potential and, critically, the practical challenges of applying two leading classes of hybrid algorithms to satellite network routing. The first class includes variational quantum optimizers, such as the Variational Quantum Eigensolver (VQE) \cite{peruzzo2014variational} and the Quantum Approximate Optimization Algorithm (QAOA) \cite{farhi2014quantum}. These algorithms reframe an optimization task as a search for the ground state of a corresponding Ising Hamiltonian, making them conceptually well-suited for static, offline route planning. The second class is Quantum-enhanced Reinforcement Learning (QRL) \cite{jerbi2021variational}, which equips a reinforcement learning agent with a Parameterized Quantum Circuit (PQC) as its policy-making core. This approach is naturally tailored for sequential decision-making in the dynamic environments that characterize satellite networks.

The primary contribution of this work is to provide a realistic benchmark of these quantum approaches, focusing on the practical hurdles encountered during implementation. In a field often characterized by hype, this type of critical assessment is crucial for guiding research efforts toward genuinely promising avenues. To achieve this, we formally model the dynamic routing problem and decompose it into two distinct sub-problems: a static shortest path problem for offline planning and a dynamic routing decision problem for online adaptation. We then implement and apply VQE, QAOA, and a QRL agent to these respective problems, detailing the process of mapping them to appropriate quantum representations such as Ising Hamiltonians and a PQC-based policy. Through extensive classical simulation of quantum circuits, our detailed performance analysis, contrary to demonstrating a quantum advantage, reveals the significant difficulties faced by these algorithms. Our analysis highlights critical failure modes, including the convergence of static optimizers to invalid solutions and the failure of the QRL agent to learn an effective policy. By carefully dissecting these informative "negative" results, we provide valuable insights into the underlying challenges of intractable optimization landscapes and learning instability, ultimately concluding with concrete future research directions required to overcome these identified obstacles.

The remainder of this paper is structured as follows. Section~\ref{sec:background} provides the necessary background on the satellite routing problem and the quantum algorithms under study. Section~\ref{sec:formulation} details our formal problem definition and the mapping to quantum models. Section~\ref{sec:methodology} describes the implementation of our algorithms and simulation environment. Section~\ref{sec:experiments} outlines the experimental setup. The results of our comparative analysis are presented in Section~\ref{sec:results}. Finally, in Section~\ref{sec:discussion} and Section~\ref{sec:conclusion}, we discuss the implications of our findings and conclude with suggestions for future research.

\section{Background and Preliminaries}
\label{sec:background}

This section provides the foundational concepts necessary to understand our work. We begin by providing a more detailed description of the satellite network routing problem, then introduce the hybrid quantum-classical computing paradigm that underpins our chosen algorithms, and finally provide an overview of VQE, QAOA, and QRL.

\subsection{The Satellite Network Routing Problem}
\label{sec:background_satellite}
As introduced in Section~\ref{sec:introduction}, routing in LEO satellite constellations is a challenging optimization task, fundamentally modeled as finding optimal paths on a dynamic graph $G(t)=(V, E(t))$ \cite{Zhang2023,Zhang2023b}. While the primary objective is typically to minimize end-to-end latency, a real-world formulation must contend with a variety of operational constraints that increase its complexity. For instance, each communication link possesses a finite \textbf{link capacity}, meaning routing decisions must avoid creating congestion that would introduce unpredictable queuing delays \cite{Grislain2024}. Furthermore, the satellites themselves operate under strict \textbf{resource limitations}, including power and processing capabilities, which may constrain the number of active links or the volume of traffic a single node can handle \cite{Xia2024}. Critically, to prevent routing loops and the inefficient use of resources, the chosen path must often adhere to a \textbf{path validity} constraint, such as being a simple path with no repeated nodes \cite{Grislain2024}. The inclusion of such multifaceted constraints elevates the problem from a simple shortest path calculation - which is solvable in polynomial time with classical algorithms like Dijkstra's - to a constrained combinatorial optimization problem that is, in its general form, NP-hard.

\subsection{The Hybrid Quantum-Classical Paradigm}
\label{sec:background_hybrid}
\begin{figure}[ht!]
    \centering
    \includegraphics[width=0.7\textwidth]{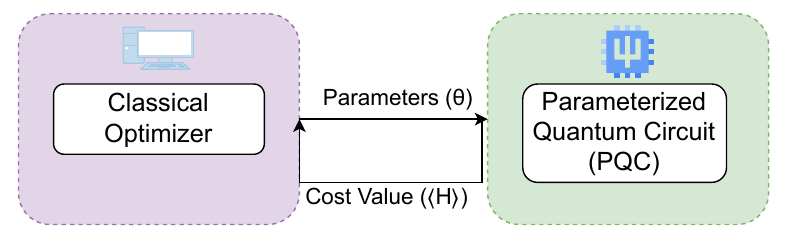} 
    \caption{The general structure of a hybrid quantum-classical variational algorithm. A classical optimizer iteratively updates the parameters of a quantum circuit to minimize a cost function evaluated on the quantum processor.}
    \label{fig:vqa_loop}
\end{figure}

The algorithms explored in this paper operate within the hybrid quantum-classical computing model, the dominant paradigm for the NISQ era \cite{DeLuca2021}. This model leverages the strengths of both classical and quantum processors by engaging them in a synergistic feedback loop, commonly known as a Variational Quantum Algorithm (VQA) \cite{Ge2022}. As illustrated in Figure~\ref{fig:vqa_loop}, this iterative process begins with the ansatz preparation step, where a classical computer provides a set of parameters, $\vec{\theta}$, to a quantum processor. The quantum device then prepares a corresponding quantum state $|\psi(\vec{\theta})\rangle$ by executing a Parameterized Quantum Circuit, or ansatz. Following this, a measurement is performed. An observable, represented by a Hamiltonian operator $\hat{H}$ that encodes the problem's cost function, is measured with respect to the prepared state, yielding an expectation value $\langle \hat{H} \rangle_{\vec{\theta}}$. This value represents the cost of the potential solution encoded by the quantum state. Finally, in the classical optimization step, this cost is returned to the classical computer, where an optimization algorithm like Adam or SPSA proposes an updated set of parameters, $\vec{\theta}'$, designed to minimize the cost. This loop repeats until the cost value converges, allowing the quantum processor to handle the classically intractable task of exploring a high-dimensional state space, while the classical processor manages the optimization search.

\subsection{Overview of Compared Algorithms}
\label{sec:background_algorithms}
\subsubsection{Variational Quantum Eigensolver (VQE)}
The VQE is a flagship VQA designed to find the lowest eigenvalue of a given Hamiltonian, which is equivalent to finding its ground state energy \cite{peruzzo2014variational}. Its operation is a direct application of the Rayleigh-Ritz variational principle, which guarantees that the expectation value of a Hamiltonian is always greater than or equal to its lowest eigenvalue: $\langle\psi|\hat{H}|\psi\rangle \ge E_0$. When applied to optimization, the task is first mapped to a problem Hamiltonian $H_P$ whose ground state corresponds to the optimal solution. 

VQE then employs a PQC, or ansatz, to prepare a trial state $|\psi(\vec{\theta})\rangle$ and iteratively adjusts the parameters $\vec{\theta}$ to find the minimum possible energy $\langle H_P \rangle_{\vec{\theta}}$. The power and flexibility of VQE lie in the choice of ansatz, which can range from physically-inspired circuits to more generic, hardware-efficient ansatze designed to be robust against noise on near-term devices.

\subsubsection{Quantum Approximate Optimization Algorithm (QAOA)}
QAOA is another prominent VQA, but one that is specifically tailored for combinatorial optimization problems \cite{farhi2014quantum}. Unlike the more general VQE, QAOA employs a problem-specific ansatz structure inspired by the principles of adiabatic quantum computation. This ansatz is constructed by the repeated application of two distinct operators. 

The first is the \textbf{Cost Hamiltonian}, $H_C$, which is diagonal in the computational basis and directly encodes the problem's objective function. The second is the \textbf{Mixer Hamiltonian}, $H_M$, typically a sum of Pauli-X operators ($H_M = \sum_i X_i$), which facilitates transitions between different potential solutions in the computational basis. Starting from an equal superposition of all states, $|+\rangle^{\otimes n}$, the QAOA circuit of depth $p$ applies the corresponding unitary operators, $U(H_C, \gamma) = e^{-i\gamma H_C}$ and $U(H_M, \beta) = e^{-i\beta H_M}$, in alternation for $p$ layers. A classical optimizer then searches for the optimal set of $2p$ variational parameters $(\vec{\gamma}^*, \vec{\beta}^*)$ that minimizes the final expectation value of the cost Hamiltonian.
\begin{equation}
    |\psi(\vec{\gamma}, \vec{\beta})\rangle = U(H_M, \beta_p) U(H_C, \gamma_p) \cdots U(H_M, \beta_1) U(H_C, \gamma_1) |+\rangle^{\otimes n}
\end{equation}

\subsubsection{Quantum-enhanced Reinforcement Learning (QRL)}
Reinforcement Learning (RL) is a machine learning paradigm where an agent learns to make optimal decisions by interacting with an environment, receiving rewards or penalties for its actions. In QRL, this framework is augmented by replacing a component of the classical RL agent with a quantum model, typically a PQC acting as a function approximator \cite{jerbi2021variational}. This work focuses on a foundational QRL architecture: a policy-based agent where the policy, $\pi(a|s)$, is directly represented by a PQC acting as a Quantum Neural Network (QNN). 

The agent learns through a well-defined cycle: first, the current state of the environment, $s$, is encoded into the PQC's input gates. The PQC, governed by its set of trainable parameters $\vec{\theta}$, is then executed. The measurement outcomes are post-processed to produce a probability distribution over all possible actions, from which an action is sampled and executed in the environment. Finally, based on the resulting reward, the agent's parameters $\vec{\theta}$ are updated using a policy gradient algorithm. The gradients required for this update can be calculated directly on quantum hardware using techniques like the parameter-shift rule \cite{schuld2019evaluating}, allowing the agent to iteratively refine its policy. This framework is inherently designed to learn complex, adaptive behaviors, making it a natural fit for dynamic environments like satellite networks.

\section{Problem Formulation and Mapping to Quantum Models}
\label{sec:formulation}

To apply quantum algorithms to the complex task of satellite network routing, we must first abstract the problem into a formal mathematical language. This section details this crucial translation process. We begin by defining the static routing problem as a constrained combinatorial optimization task. We then describe its transformation into a Quadratic Unconstrained Binary Optimization (QUBO) model suitable for VQE and QAOA, and finally, we reformulate the dynamic routing problem as a Markov Decision Process for our QRL agent.

\subsection{Static Routing as a Constrained Combinatorial Optimization Problem}
For the static routing scenarios, which model the task of offline route planning on a network snapshot, we consider a fixed, undirected graph $G = (V, E)$. Here, $V$ is the set of $N=|V|$ nodes (satellites or ground stations), and $E$ is the set of available communication links. Each edge $(i, j) \in E$ is assigned a positive weight $w_{ij} > 0$, representing the communication latency. The objective is to identify a simple (loop-free) path from a designated source node $s \in V$ to a destination node $d \in V$.

To express this problem algebraically, we introduce a matrix of binary variables, $x_{i,k}$, which forms the basis of our encoding. A variable $x_{i,k}$ is defined to take a value of 1 if node $i$ is the $k$-th node in the path, and 0 otherwise. Formally:
\begin{equation}
    x_{i,k} \in \{0, 1\} \quad \forall i \in V, \forall k \in \{0, \dots, N-1\}
\end{equation}
This encoding requires a total of $N^2$ binary variables to represent any possible path of up to length $N-1$. While more qubit-efficient logarithmic encodings exist, we select this straightforward $N^2$ representation to establish a clear baseline for the performance of VQAs on penalty-based constrained optimization problems, deferring the investigation of more complex encodings to future work. The primary goal is to minimize the total path latency. This objective can be formulated as a quadratic cost function, $C(\mathbf{x})$, which sums the weights of consecutive edges in the path:
\begin{equation}
    \text{minimize} \quad C(\mathbf{x}) = \sum_{(i,j) \in E} w_{ij} \sum_{k=0}^{N-2} x_{i,k} x_{j,k+1}
\end{equation}
where $\mathbf{x}$ denotes the vector of all $N^2$ binary variables.

However, minimizing this cost function alone is insufficient, as an arbitrary assignment of binary values to $\mathbf{x}$ will likely not represent a valid physical path. Therefore, we must impose a set of linear constraints. A \textbf{start point constraint} enforces that the path originates at the source node $s$ at the first position ($k=0$). Furthermore, a \textbf{position uniqueness constraint} dictates that each position in the path can be occupied by exactly one node. Finally, to ensure the path is simple, a \textbf{node uniqueness constraint} ensures that each node can be visited at most once. These constraints are expressed mathematically as:
\begin{align}
    x_{s,0} &= 1 \label{eq:constraint_start} \\
    \sum_{i \in V} x_{i,k} &= 1 \quad \forall k \in \{0, \dots, N-1\} \label{eq:constraint_pos_unique} \\
    \sum_{k=0}^{N-1} x_{i,k} &\le 1 \quad \forall i \in V \label{eq:constraint_node_unique}
\end{align}

\subsection{Mapping to a QUBO Model for VQE and QAOA}
The constrained optimization problem defined above cannot be solved directly by standard VQE or QAOA, which are designed to find the ground state of unconstrained physical systems. The first step, therefore, is to transform it into an unconstrained problem. This is achieved by reformulating it as a QUBO problem. The QUBO formalism is a cornerstone of both classical and quantum annealing, as well as variational algorithms, and seeks to minimize a function of the form:
\begin{equation}
    f(\mathbf{x}) = \mathbf{x}^T Q \mathbf{x} = \sum_{i,j} Q_{ij} x_i x_j
\end{equation}
where $\mathbf{x}$ is a vector of binary variables and $Q$ is a real-valued square matrix encoding the problem's structure.

The transformation involves integrating the constraints (Equations \ref{eq:constraint_start}-\ref{eq:constraint_node_unique}) into the objective function as quadratic penalty terms. Each penalty term is constructed to have a value of zero when its corresponding constraint is satisfied and a large positive value when it is violated. The complete QUBO objective function is a weighted sum of the original cost and these penalties:
\begin{equation}
    H_{\text{QUBO}}(\mathbf{x}) = C(\mathbf{x}) + P \cdot \left( \sum_{\text{constraints } j} \text{Penalty}_j(\mathbf{x}) \right)
\end{equation}
Here, $P$ is a penalty coefficient, a crucial hyperparameter that must be set large enough to dominate the cost term $C(\mathbf{x})$. Selecting an appropriate value for $P$ is a critical aspect of the formulation; a common heuristic is to set $P$ to a value greater than the maximum possible path cost in the graph. This guarantees that any single constraint violation incurs a penalty larger than the cost of the most "expensive" possible valid path, thereby ensuring that the optimizer prioritizes finding valid solutions. For instance, the penalty for the position uniqueness constraint (Equation \ref{eq:constraint_pos_unique}) can be written as $P \left( 1 - \sum_{i \in V} x_{i,k} \right)^2$.

The final step in this mapping is to convert the classical QUBO expression into a quantum-mechanical Ising Hamiltonian, the native input for our variational algorithms. This is accomplished via the standard transformation $x_i \to (1 - Z_i)/2$, where $Z_i$ is the Pauli-Z operator acting on the $i$-th qubit. This process results in an Ising problem Hamiltonian of the general form:
\begin{equation}
    H_P = \sum_{i} h_i Z_i + \sum_{i<j} J_{ij} Z_i Z_j + \text{const.}
    \label{eq:ising_hamiltonian_expanded}
\end{equation}
The coefficients $h_i$ (local fields) and $J_{ij}$ (couplings) are derived directly from the entries of the QUBO matrix $Q$. The ground state of this Hamiltonian $H_P$ corresponds to the binary string $\mathbf{x}$ that minimizes the QUBO function, thus encoding the optimal shortest path.

\subsection{Dynamic Routing as a Markov Decision Process for QRL}
For the dynamic routing scenario, where the network $G(t)$ evolves over time, a static optimization approach is fundamentally unsuitable. We therefore shift our paradigm from optimization to sequential decision-making. This problem is naturally modeled as a Markov Decision Process (MDP), the mathematical foundation of reinforcement learning. An MDP is formally defined by the tuple $(\mathcal{S}, \mathcal{A}, \mathcal{P}, \mathcal{R})$.

The \textbf{state space} $\mathcal{S}$ encompasses all information necessary for the agent to make an informed decision. We define a state $s_t \in \mathcal{S}$ as a tuple $s_t = (u, d, A(t))$, where $u$ is the packet's current location, $d$ is its final destination, and $A(t)$ is the network's adjacency matrix at time $t$. This representation provides the agent with full observability of the immediate network topology.

The \textbf{action space} $\mathcal{A}$ defines the set of possible decisions. At a node $u$, the action space $\mathcal{A}_u$ is the set of all neighboring nodes. An action $a_t \in \mathcal{A}_u$ is the selection of one of these neighbors as the next hop for the packet.

The \textbf{transition probability} $\mathcal{P}(s'|s, a)$ governs the environment's dynamics. In our simulation, the environment is deterministic with respect to the agent's action (the packet moves to the chosen node) but stochastic with respect to topology changes, which occur at fixed intervals independent of the agent's actions.

Finally, the \textbf{reward function} $\mathcal{R}(s, a)$ provides the learning signal. It is carefully designed to guide the agent towards the desired behavior of finding low-latency paths quickly. We define it as a multi-component function:
\begin{equation}
    R(s_t, a_t=v) = 
    \begin{cases} 
        R_{\text{dest}} & \text{if } v = d \\
        -w_{uv}(t) & \text{if } (u,v) \in E(t) \text{ and } v \neq d \\
        R_{\text{invalid}} & \text{if } (u,v) \notin E(t)
    \end{cases}
\end{equation}
Here, $R_{\text{dest}}$ is a large positive reward for reaching the destination, $-w_{uv}(t)$ is a negative reward equal to the latency of the chosen link, and $R_{\text{invalid}}$ is a large negative reward to penalize invalid actions. The agent's objective is to learn a policy $\pi(a|s)$ that maximizes the long-term cumulative reward. In our QRL framework, this policy is parameterized by a PQC.

\section{Methodology and Implementation Details}
\label{sec:methodology}

This section provides a comprehensive description of our implementation methodology. We detail the specific construction of each quantum algorithm and describe the classical simulation environment used to conduct our experiments. All quantum components are implemented using the PennyLane quantum computing library \cite{bergholm2018pennylane}, integrated with PyTorch for automatic differentiation.

\subsection{VQE Implementation for the Shortest Path Problem}
Our VQE solver is designed to find the ground state of the Ising Hamiltonian $H_P$, which encodes the static shortest path problem. The overall process follows the standard variational quantum eigensolver loop, as detailed in Algorithm~\ref{alg:vqe}.

\begin{algorithm}[ht!]
\caption{Variational Quantum Eigensolver (VQE) for Shortest Path}
\footnotesize
\label{alg:vqe}
\begin{algorithmic}[1]
\State \textbf{Input:} Graph $G$, source $s$, destination $d$, penalty $P$, layers $L$, steps $T$, learning rate $\alpha$
\State \textbf{Output:} Optimal path approximation $\text{path}^*$

\State \Function{VQE\_Solver}{$G, s, d, P, L, T, \alpha$}
    \State Build QUBO matrix $Q$ from $G, s, d, P$
    \State Convert $Q$ to Ising Hamiltonian $H_P$
    \State Initialize variational parameters $\vec{\theta}$ randomly for an $L$-layer ansatz
    \State Initialize Adam optimizer with parameters $\vec{\theta}$ and learning rate $\alpha$
    
    \For{$t = 1 \to T$}
        \State Evaluate cost on QPU: $E(\vec{\theta}) \gets \langle\psi(\vec{\theta})|H_P|\psi(\vec{\theta})\rangle$
        \State Compute gradients on QPU: $\vec{g} \gets \nabla_{\vec{\theta}} E(\vec{\theta})$
        \State Update parameters via optimizer: $\vec{\theta} \gets \text{AdamUpdate}(\vec{\theta}, \vec{g}, \alpha)$
    \EndFor
    
    \State Let $\vec{\theta}^*$ be the final optimized parameters
    \State Prepare final state $|\psi(\vec{\theta}^*)\rangle$ and measure in the computational basis
    \State Let $\mathbf{b}^*$ be the most frequently measured bitstring
    \State Decode $\mathbf{b}^*$ into a path representation $\text{path}^*$
    \State \textbf{return} $\text{path}^*$
\EndFunction
\end{algorithmic}
\end{algorithm}

The core of the VQE is the \textbf{ansatz}, for which we employ the \texttt{BasicEntanglerLayers} template from PennyLane. This hardware-efficient circuit of depth $L$ prepares the trial state $|\psi(\vec{\theta})\rangle$. The \textbf{classical optimizer} is responsible for updating the parameters $\vec{\theta}$ to minimize the energy expectation value $E(\vec{\theta}) = \langle\psi(\vec{\theta})|H_P|\psi(\vec{\theta})\rangle$. We utilize the Adam optimizer \cite{kingma2014adam} for this task, with gradients provided by PennyLane's implementation of the parameter-shift rule.

A critical implementation detail is the selection of the penalty coefficient $P$ for the QUBO formulation. Following the heuristic introduced in Section~\ref{sec:formulation}, we first calculated the maximum possible cost of any simple path in the experimental graph. The value of $P$ was then set to be significantly larger than this maximum cost. This ensures that any solution candidate that violates a constraint receives a total penalty that decisively outweighs the cost of even the most expensive valid path, thereby rigorously enforcing the problem's constraints within the optimization landscape.

\subsection{QAOA Implementation for the Max-Cut Problem}
For the Max-Cut control experiment, we implemented the Quantum Approximate Optimization Algorithm. The \textbf{Hamiltonians} - cost ($H_C$) and mixer ($H_M$) - were generated using standard definitions for Max-Cut. The QAOA process is outlined in Algorithm~\ref{alg:qaoa}. The key difference from VQE lies in its structured, problem-inspired ansatz, which consists of $p$ alternating layers of cost and mixer unitary evolutions. The $2p$ variational parameters $(\vec{\gamma}, \vec{\beta})$ are again optimized using the Adam optimizer.

\begin{algorithm}[ht!]
\caption{Quantum Approximate Optimization Algorithm (QAOA) for Max-Cut}
\footnotesize
\label{alg:qaoa}
\begin{algorithmic}[1]
\State \textbf{Input:} Graph $G$, layers $p$, steps $T$, learning rate $\alpha$
\State \textbf{Output:} Optimal node partition approximation $\mathbf{b}^*$

\State \Function{QAOA\_Solver}{$G, p, T, \alpha$}
    \State Build cost Hamiltonian $H_C$ and mixer Hamiltonian $H_M$ from $G$
    \State Initialize variational parameters $(\vec{\gamma}, \vec{\beta})$ randomly
    \State Initialize Adam optimizer with parameters $(\vec{\gamma}, \vec{\beta})$ and learning rate $\alpha$
    
    \For{$t = 1 \to T$}
        \State Define ansatz state $|\psi(\vec{\gamma}, \vec{\beta})\rangle = [ \prod_{k=1}^{p} e^{-i\beta_k H_M} e^{-i\gamma_k H_C} ] |+\rangle^{\otimes n}$
        \State Evaluate cost on QPU: $E(\vec{\gamma}, \vec{\beta}) \gets \langle\psi|H_C|\psi\rangle$
        \State Compute gradients on QPU: $(\vec{g}_{\gamma}, \vec{g}_{\beta}) \gets \nabla_{(\vec{\gamma}, \vec{\beta})} E(\vec{\gamma}, \vec{\beta})$
        \State Update parameters via optimizer
    \EndFor
    
    \State Let $(\vec{\gamma}^*, \vec{\beta}^*)$ be the final optimized parameters
    \State Prepare final state $|\psi(\vec{\gamma}^*, \vec{\beta}^*)\rangle$ and measure
    \State Let $\mathbf{b}^*$ be the most frequently measured bitstring
    \State \textbf{return} $\mathbf{b}^*$
\EndFunction
\end{algorithmic}
\end{algorithm}

\subsection{QRL Agent Implementation}
Our implementation is based on the REINFORCE algorithm, which we selected as a direct quantum analogue of a foundational classical policy gradient method. The goal of this choice is to establish a performance baseline for a straightforward QRL approach. We acknowledge that classical REINFORCE is well-known for its high gradient variance and sample inefficiency, limitations that are expected to persist in its quantum counterpart. By benchmarking this foundational agent, we can robustly assess the core challenges of applying QRL to this domain, which in turn motivates the exploration of more advanced algorithms discussed in Section~\ref{sec:discussion}.

The agent's policy $\pi_{\vec{\theta}}(a|s)$ is represented by a PQC, and it learns by interacting with the dynamic network environment over many episodes, as detailed in Algorithm~\ref{alg:qrl}. The \textbf{PQC Architecture} uses $\lceil \log_2(N) \rceil$ qubits and consists of a state-encoding layer followed by a trainable variational circuit. The \textbf{training process} involves collecting a trajectory of experiences within an episode. At the end of the episode, the discounted returns are calculated, and this information is used to compute a stochastic policy gradient. The gradients are estimated using the parameter-shift rule and fed into the Adam optimizer to update the PQC's parameters $\vec{\theta}$.

\begin{algorithm}[ht!]
\footnotesize
\caption{Quantum Reinforcement Learning (QRL) with REINFORCE}
\label{alg:qrl}
\begin{algorithmic}[1]
\State \textbf{Input:} Dynamic Environment \texttt{Env}, training episodes $M$
\State \textbf{Initialize:} PQC policy $\pi_{\vec{\theta}}(a|s)$ with random parameters $\vec{\theta}$
\State \textbf{Initialize:} Optimizer (e.g., Adam) for $\vec{\theta}$

\For{episode = $1 \to M$}
    \State Initialize trajectory storage: $\mathcal{T} = []$
    \State Get initial state from environment: $s_0 \gets \texttt{Env.reset()}$
    \State $t \gets 0$
    
    \While{episode is not done}
        \State Get action probabilities from PQC: $P(A|s_t) \gets \pi_{\vec{\theta}}(A|s_t)$
        \State Sample action: $a_t \sim P(A|s_t)$
        \State Execute action $a_t$ in \texttt{Env}, receive reward $r_{t+1}$ and next state $s_{t+1}$
        \State Store transition in trajectory: $\mathcal{T} \gets \mathcal{T} \cup \{(s_t, a_t, r_{t+1})\}$
        \State $s_t \gets s_{t+1}$, $t \gets t+1$
    \EndWhile
    
    \State Initialize gradient estimate $\Delta\vec{\theta} \gets 0$
    \For{each step $t=0, \dots, T-1$ in trajectory $\mathcal{T}$}
        \State Calculate discounted return: $G_t \gets \sum_{k=t}^{T-1} \gamma^{k-t} r_{k+1}$
        \State Compute policy gradient term: $\vec{g}_t \gets G_t \nabla_{\vec{\theta}} \log \pi_{\vec{\theta}}(a_t|s_t)$
        \State Accumulate gradient: $\Delta\vec{\theta} \gets \Delta\vec{\theta} + \vec{g}_t$
    \EndFor
    
    \State Update policy parameters: $\vec{\theta} \gets \text{OptimizerUpdate}(\vec{\theta}, \Delta\vec{\theta})$
\EndFor
\end{algorithmic}
\end{algorithm}

\subsection{Simulation Environment}
All experiments were conducted on a classical computer simulating an ideal, noise-free quantum processor.
\begin{itemize}
    \item \textbf{Quantum Backend:} We utilized PennyLane's high-performance statevector simulator, \texttt{default.qubit}. This allows us to benchmark the ideal performance of the algorithms, a necessary step to distinguish fundamental algorithmic bottlenecks from hardware-induced errors. The potential impact of noise is further considered in Section~\ref{sec:discussion}.
    \item \textbf{Network Simulation:} The satellite network scenarios were managed by a custom Python environment built upon the NetworkX library \cite{hagberg2008exploring}.
\end{itemize}

\section{Experiments and Setup}
\label{sec:experiments}

To systematically evaluate the performance and inherent challenges of the selected quantum algorithms, we designed a series of three distinct experiments. This series was structured to progress logically from a well-understood canonical problem to more complex static and dynamic routing tasks, allowing us to isolate and analyze the behavior of each algorithm under different conditions.

\subsection{Scenario A: Canonical Max-Cut Problem as a Control}
Our first scenario serves as a crucial control experiment to validate our algorithmic implementations, particularly for VQE and QAOA. For this purpose, we selected the Max-Cut problem, a canonical task in combinatorial optimization. Max-Cut is advantageous as a benchmark because its Hamiltonian structure is unconstrained and well-understood, simplifying the analysis of algorithmic performance. The specific instance involves a 5-node, 7-edge undirected graph. The task is to partition the five nodes into two disjoint sets in a way that maximizes the number of edges connecting nodes across the two sets. The primary purpose of this scenario is to establish a baseline for our solvers. By testing them on a problem where success is expected (especially for QAOA), we can ensure that any failures observed in more complex subsequent scenarios are attributable to the intrinsic difficulty of those problems, rather than to fundamental errors in our implementation.

\subsection{Scenario B: Static Shortest Path Problem}
This second scenario constitutes our primary test case for applying static quantum optimization algorithms to a constrained routing problem, simulating an offline planning task. We constructed a 4-node, 5-edge undirected graph with manually assigned integer weights, for which the optimal shortest path is known and can be verified by classical means (the path $0 \to 1 \to 2 \to 3$ with a total cost of 11). The central task for the VQE and QAOA algorithms is to find this specific path, starting from node 0 and ending at node 3. This problem is significantly more challenging than Max-Cut due to the necessity of satisfying multiple hard constraints, as detailed in Section~\ref{sec:formulation}. The purpose of this scenario is to rigorously assess whether VQE and QAOA, utilizing the resource-intensive $N^2$-qubit encoding scheme (resulting in a 16-qubit system), can successfully navigate the complex, penalty-laden optimization landscape to find a valid and optimal solution under ideal, noise-free conditions.

\subsection{Scenario C: Dynamic Routing Environment Simulating a LEO Constellation}
This scenario directly addresses the core challenge articulated in this paper's title: routing in a dynamic satellite network. It is specifically designed to test the learning and adaptation capabilities of the QRL agent in a non-stationary environment that abstracts the key dynamic properties of an LEO constellation.

For this purpose, we first generate an 8-node Barabási-Albert graph. This model is chosen because it naturally creates scale-free networks with a few highly connected "hub" nodes, a topology characteristic of robust communication networks and analogous to satellites with multiple active ISLs. The edge weights are integers chosen uniformly from [1, 10], representing heterogeneous latencies.

The crucial aspect of this scenario is its \textbf{dynamism}, which simulates two primary phenomena in LEO networks. First, the high orbital velocity of satellites means that ISLs are constantly being established with new neighbors while connections to receding satellites are terminated. Second, link quality can degrade or fail temporarily. We model these effects with a simple yet effective mechanism: at a fixed interval of 10 steps within each training episode, a random existing edge is removed from the graph (simulating a link termination or failure), and a new random edge is added between two previously unconnected nodes (simulating the establishment of a new line-of-sight link).

The task for the QRL agent within this evolving environment is to learn a generalizable, state-aware policy. In each episode, the agent is presented with a randomly chosen source-destination pair and must route a packet to its destination within a maximum of 25 steps. The purpose of this scenario is twofold: first, to evaluate whether the foundational QRL agent, based on the REINFORCE algorithm, can learn an effective routing policy that outperforms a naive, random-choice baseline in a non-stationary setting. Second, and more importantly, it aims to demonstrate the fundamental inadequacy of static optimization methods, whose pre-calculated "optimal" paths would be rendered obsolete by the first topological change.

\subsection{Performance Metrics}
To ensure a fair and comprehensive comparison, we defined distinct sets of performance metrics tailored to the nature of each algorithmic class.

For the \textbf{static optimization algorithms} (VQE and QAOA) in Scenarios A and B, our primary metric is \textbf{solution validity}. Since finding a valid solution that satisfies all constraints is a non-trivial task in itself, this serves as the first-pass criterion for success. If a valid solution is found, we then evaluate its quality using the \textbf{approximation ratio}, defined as the ratio of the found cost to the known optimal cost. An approximation ratio of 1.0 signifies that the optimal solution was found. Additionally, we analyze the \textbf{convergence behavior} of the optimizers by plotting the energy (cost function value) versus the number of optimization steps, providing insight into the dynamics of the learning process.

For the \textbf{dynamic learning agent} (QRL) in Scenario C, performance is measured over time. The two key metrics are the \textbf{success rate} and the \textbf{average reward per episode}. The success rate is the percentage of episodes in which the agent successfully routes the packet to its destination within the allotted time. These metrics are tracked over a moving window of 100 episodes to smooth out short-term fluctuations and produce clear learning curves, which visualize the agent's performance improvement (or lack thereof) throughout the training process.

\section{Results and Analysis}
\label{sec:results}

This section presents the empirical results from our simulations. The findings from all three scenarios consistently highlight the significant practical challenges faced by the variational quantum algorithms when applied to network optimization problems, providing a realistic assessment of their current capabilities.

\subsection{Static Optimizers Fail on Constrained Routing Task}
Our first key finding is the categorical failure of both VQE and QAOA to solve the static shortest path problem (Scenario B). This outcome is particularly insightful when contrasted with their performance on the simpler, unconstrained Max-Cut problem (Scenario A), which serves as our control experiment.

The results for both static scenarios are summarized in Table~\ref{tab:static_results}. For the 5-node Max-Cut problem, the QAOA implementation successfully identified the optimal node partition, achieving an approximation ratio of 1.0. This successful execution on a canonical benchmark validates the correctness of our general workflow, including the optimizer and result interpretation logic. In contrast, the VQE solver, using a generic hardware-efficient ansatz, failed to find the optimal solution and instead converged to a trivial, invalid state (a partition where all nodes are in the same set).

\begin{table}[ht!]
\centering
\caption{Summary of Static Optimization Results. The approximation ratio is considered invalid if the algorithm fails to produce a valid solution that satisfies all problem constraints.}
\label{tab:static_results}
\begin{tabular}{@{}llcc@{}}
\toprule
Problem & Algorithm & Solution Validity & Approx. Ratio \\ \midrule
\textbf{5-Node Max-Cut} & VQE & Invalid & 0.0 \\
(Scenario A) & QAOA & \textbf{Valid (Optimal)} & \textbf{1.0} \\ \addlinespace
\textbf{4-Node Shortest Path} & VQE & Invalid & Invalid \\
(Scenario B) & QAOA & Invalid & Invalid \\ \bottomrule
\end{tabular}
\end{table}

\begin{figure}[ht!]
    \centering
    \begin{subfigure}{0.48\textwidth}
        \centering
        \includegraphics[width=\linewidth]{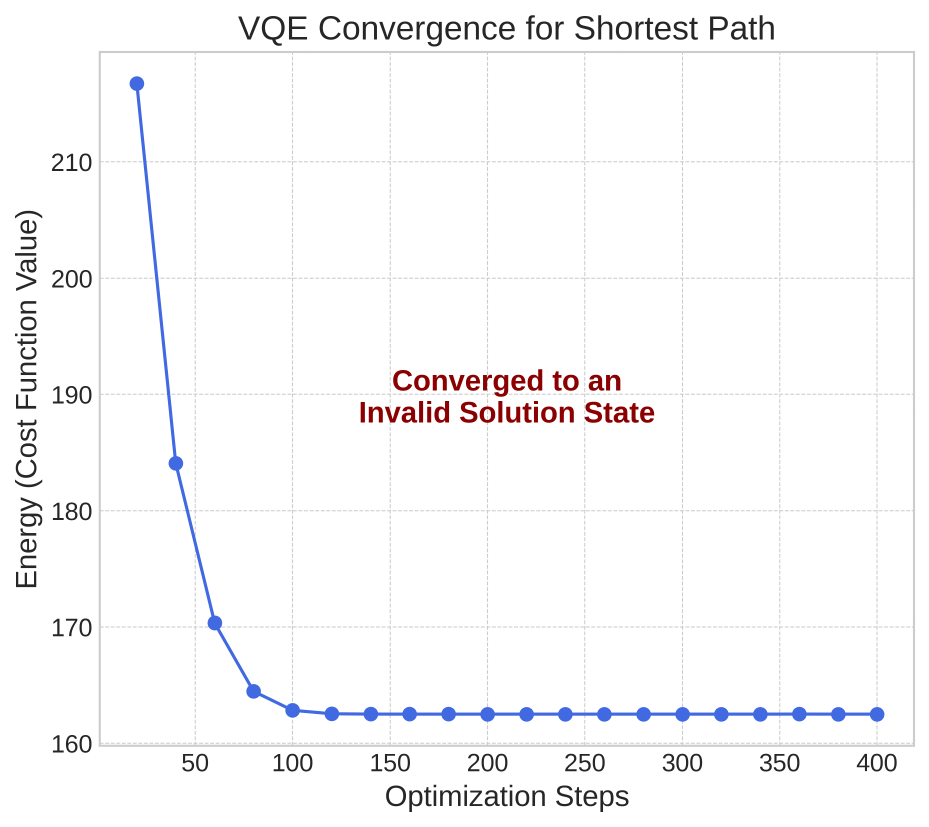}
        \caption{VQE Convergence for Shortest Path}
        \label{fig:vqe_sp_conv}
    \end{subfigure}
    \hfill
    \begin{subfigure}{0.48\textwidth}
        \centering
        \includegraphics[width=\linewidth]{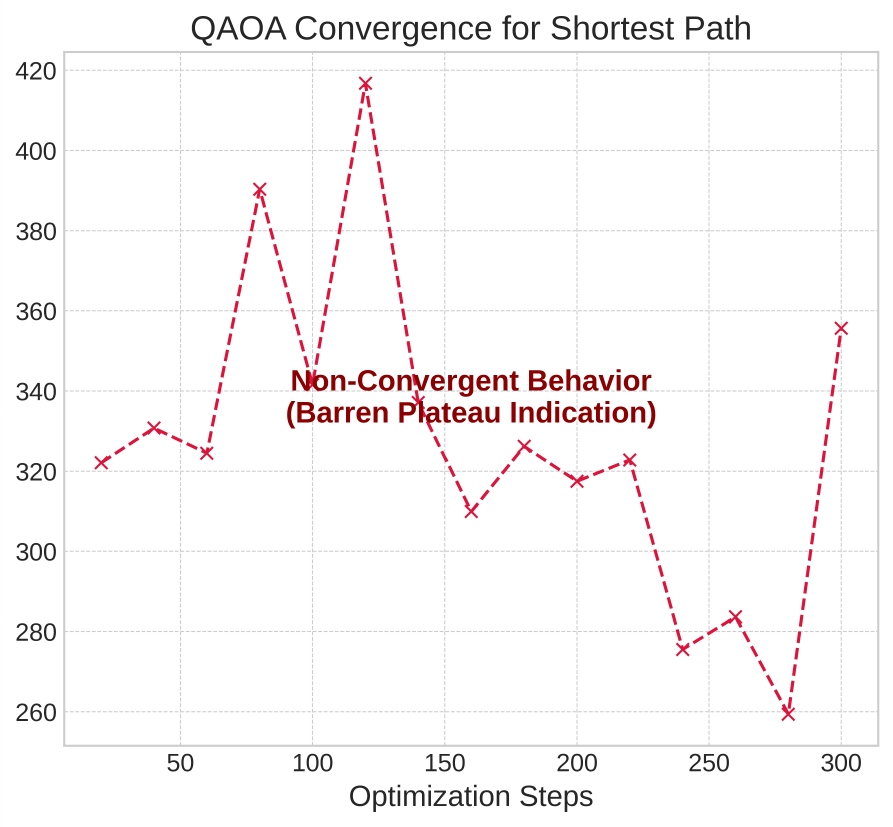}
        \caption{QAOA Convergence for Shortest Path}
        \label{fig:qaoa_sp_conv}
    \end{subfigure}
    \caption{Convergence behavior for the 4-node (16-qubit) shortest path problem. (a) VQE converges smoothly to a stable energy value, but the corresponding state is invalid. (b) QAOA exhibits erratic, non-convergent behavior, indicative of a barren plateau.}
    \label{fig:static_sp_results}
\end{figure}

When faced with the more complex, constrained shortest path problem, both algorithms failed to produce a valid path. Figure~\ref{fig:static_sp_results} provides a deeper insight into their distinct failure modes. The VQE optimization process, shown in Figure~\ref{fig:vqe_sp_conv}, exhibits smooth convergence. The energy of the variational state decreases steadily and stabilizes at a low value. However, upon inspection, the computational basis state corresponding to this energy minimum encodes an invalid path, violating the fundamental start/end point constraints. This indicates that the optimization process became irreversibly trapped in a deep local minimum within the vast 16-qubit Hilbert space that, while energetically favorable, does not correspond to a valid solution to the original problem.

More strikingly, the QAOA optimization process, depicted in Figure~\ref{fig:qaoa_sp_conv}, failed to converge at all. The energy value fluctuates erratically throughout the optimization, showing no discernible downward trend. This behavior is a strong indicator of an underlying barren plateau phenomenon \cite{mcclean2018barren}. The complex, penalty-laden Hamiltonian required for the constrained problem likely creates an optimization landscape where gradients vanish exponentially across most of the parameter space, rendering gradient-based optimizers like Adam ineffective.

\subsection{QRL Agent Fails to Learn an Effective Dynamic Policy}
Our second key finding is that the foundational QRL agent also failed to learn a useful policy in the dynamic network environment of Scenario C. Despite extensive training over 3000 episodes and substantial hyperparameter tuning (including adjustments to learning rate and PQC depth), the agent's performance did not demonstrate any sustained improvement and remained comparable to a simple random-choice baseline.

\begin{figure}
    \centering
    \includegraphics[width=0.8\textwidth]{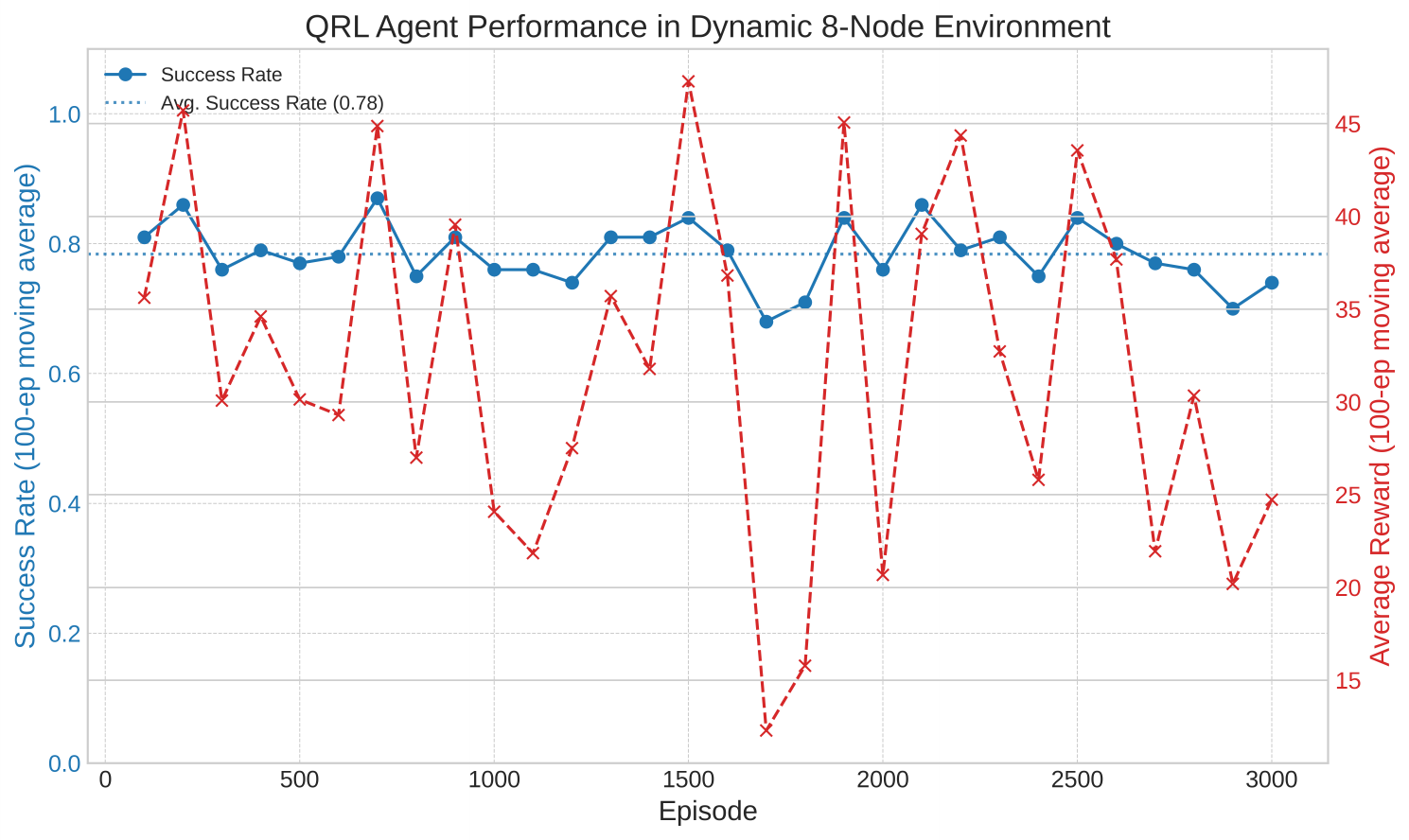}
    \caption{Learning curve for the QRL agent in the dynamic 8-node environment. Both the success rate and the average reward (calculated over a 100-episode moving window) fail to show a consistent upward trend, indicating a lack of effective learning over 3000 episodes.}
    \label{fig:qrl_results}
\end{figure}

Figure~\ref{fig:qrl_results} plots the key performance metrics throughout the training process. The success rate (blue line) fluctuates between 70\% and 87\% but shows no clear learning trend, ultimately ending in the same performance regime as a naive random agent. Similarly, the average reward per episode (red line), which should increase as the agent learns to find shorter paths, remains stagnant and highly variant. This result suggests that the vanilla policy gradient algorithm (REINFORCE) is insufficient for this task. The high variance of the gradient estimates, compounded by the non-stationarity of the environment (i.e., the constantly changing network topology), likely provides an unstable and unreliable learning signal. Consequently, the optimizer is unable to consistently update the PQC parameters in a direction that leads to a better policy, even though the PQC itself may possess the expressive power to represent one.

\section{Discussion and Future Work}
\label{sec:discussion}

The collective results of our experiments paint a sober but crucial picture of the current state of near-term quantum algorithms for complex, constrained problems like network routing. Rather than demonstrating a quantum advantage, our findings highlight fundamental challenges in both the formulation and optimization of these algorithms. This section analyzes the root causes of these challenges and proposes several promising research avenues to address them.

\subsection{Analysis of Algorithmic Challenges}
The failure of VQE and QAOA on the static shortest path problem can be primarily attributed to the immense complexity of the problem's Hamiltonian representation. The $N^2$-qubit encoding scheme, while straightforward to formulate, results in an exponential increase in the Hilbert space dimension that the variational algorithm must search. This vast space, combined with the large penalty terms required to enforce path constraints, creates an optimization landscape that is likely riddled with deep local minima corresponding to invalid solutions. Our results show that a generic, hardware-efficient ansatz, as used in our VQE, is not sufficiently structured to navigate this landscape and consistently becomes trapped. Perhaps more surprisingly, even the problem-inspired ansatz of QAOA was unable to find a path toward convergence, suggesting that the intricate interplay between the cost and mixer Hamiltonians in this high-penalty regime is highly non-trivial to optimize and prone to issues like barren plateaus.

For the QRL agent, the observed failure to learn points to the inherent instability of the foundational REINFORCE algorithm. Policy gradient methods are notoriously sample-inefficient and suffer from high variance in their gradient estimates. In our dynamic routing environment, this issue is severely exacerbated. The non-stationarity of the network topology means that the reward landscape is constantly shifting, providing a noisy and unreliable learning signal to the agent. While the PQC offers a potentially powerful and expressive policy representation, its theoretical advantages cannot be realized if the learning algorithm itself is unstable. The agent is unable to distinguish between a bad outcome due to a poor policy and a bad outcome due to an unlucky environmental change, preventing effective credit assignment and meaningful parameter updates.

It is critical to underscore that our findings were obtained using an ideal, noise-free quantum simulator. This deliberate choice allowed us to isolate the algorithmic and formulation-based challenges discussed above. The failures we observed are therefore not attributable to hardware imperfections but are fundamental to the algorithms themselves in this problem context. This finding makes our conclusions stronger: if these foundational methods struggle on complex, penalty-laden landscapes and with unstable learning signals even under perfect conditions, their performance would inevitably be further degraded on real NISQ-era hardware. The presence of gate errors, qubit decoherence, and measurement noise would likely exacerbate the difficulty of navigating optimization landscapes and extracting reliable gradients, presenting another significant layer of difficulty on the path to achieving a practical quantum advantage.

\subsection{Future Directions and Proposed Solutions}
Our findings, though cautionary, are not a verdict against the potential of quantum computing for this domain. Instead, they illuminate a clear and necessary path forward, shifting the focus from simple benchmarking to addressing fundamental algorithmic and representational challenges. We propose that future research prioritize the following areas.

First, for static optimization, there is a critical need for \textbf{more efficient problem encodings}. The primary bottleneck identified in our work was the qubit-intensive $N^2$ representation. Future work should investigate more compact schemes. One promising direction is the development of logarithmic encodings, which could potentially represent paths on $O(\log N)$ or $O(N \log N)$ qubits, drastically reducing the resource requirements and potentially simplifying the optimization landscape.

Second, to make variational optimizers viable for such problems, research into \textbf{advanced training techniques and ansatz designs} is essential. To combat the barren plateau problem observed with QAOA, techniques such as layer-wise training, parameter initialization strategies based on classical heuristics, or the development of ansatze with locality structures that reflect the graph's topology could be explored.

The third and arguably most promising direction is the development of more \textbf{stable and sample-efficient Quantum Reinforcement Learning algorithms}. The limitations of REINFORCE are well-known in the classical domain, and its quantum counterpart is no exception. We strongly propose that future work investigate the implementation of Quantum Actor-Critic (QAC) methods \cite{dunjko2018machine}. In a QAC architecture, a second model - the critic (which could be quantum or classical) - is trained to estimate the value function, $V(s)$. This provides a stable baseline for the policy gradient, allowing the agent to learn based on the "advantage" of an action ($A(s,a) = R + \gamma V(s') - V(s)$) rather than the raw, noisy reward signal. This has been shown to dramatically improve stability and performance in classical RL and represents a logical and powerful next step for QRL.

Finally, the interface between classical data and the quantum policy - the state representation - can be significantly improved. Instead of simply flattening the adjacency matrix, future work could explore \textbf{Quantum Graph Neural Networks (QGNNs)} \cite{verdon2019quantum}. QGNNs provide a natural and powerful way to encode graph-structured data into a PQC by using parameterized quantum operations that mimic the message-passing and aggregation steps of classical GNNs. This could allow the QRL agent to learn more relevant and robust features of the network topology, leading to more effective routing policies.

\section{Conclusion}
\label{sec:conclusion}

This paper presented a systematic analysis of near-term hybrid quantum-classical algorithms applied to the challenging domain of dynamic satellite network routing. Our work provides a realistic benchmark of the capabilities and - more critically - the limitations of two dominant paradigms: variational quantum optimization (VQE and QAOA) for static route planning, and quantum reinforcement learning for dynamic policy-making.

Our principal finding is that both approaches, in their foundational forms, face significant and fundamental hurdles when applied to this problem class. We demonstrated that variational optimizers struggled immensely with the constrained shortest path problem, where VQE consistently converged to invalid solutions and QAOA exhibited non-convergent behavior indicative of barren plateaus. Similarly, our QRL agent, based on the REINFORCE algorithm, was unable to learn an effective routing policy in a non-stationary environment, showing no discernible performance improvement over a random baseline.

These results lead to a crucial conclusion: achieving a practical quantum advantage in complex, real-world applications like network routing is not merely a matter of hardware scaling but is equally dependent on significant advances in the algorithms themselves. The challenges we identified - related to problem encoding, optimization landscape complexity, and learning algorithm instability - must be addressed head-on. By identifying these specific hurdles and proposing concrete future directions, such as the exploration of more stable Quantum Actor-Critic methods and efficient Quantum Graph Neural Network encodings, this work provides a valuable and pragmatic roadmap. We hope that this realistic assessment of current challenges will help guide future research toward developing the robust and scalable quantum solutions required for next-generation communication networks.

\section*{Code Availability}
\addcontentsline{toc}{section}{Code Availability} 

The complete source code for the simulation environment, implementation of the VQE, QAOA, and QRL algorithms, and the scripts required to reproduce all figures and tables in this study are openly available on GitHub at the following repository: \url{https://github.com/ailabteam/Quantum-AI-Network-Optimization}.


\bibliographystyle{plainnat} 

\end{document}